\documentclass[letterpaper,english,reprint, aps]{revtex4-1}
\usepackage{graphicx}
\usepackage{amsmath,amssymb}
\usepackage[hidelinks]{hyperref}
\usepackage{float}

\newcommand{\nocontentsline}[3]{}
\newcommand{\tocless}[2]{\bgroup\let\addcontentsline=\nocontentsline#1{#2}\egroup}
\newcommand{\bvec}[1]{{\mathbf #1}}

\newcommand{\ket}[1]{\left| #1 \right>}

\begin{document}

\title{Persistence of chirality in the Su-Schrieffer-Heeger model in the presence of on-site disorder}

\author{Myles Scollon and Malcolm P. Kennett}
\affiliation{
 Department of Physics, Simon Fraser University\\
 8888 University Drive, Burnaby, British Columbia, V5A 1S6, Canada}

\date{\today}

\begin{abstract}
We consider the effects of on-site  and hopping disorder on zero modes in the Su-Schrieffer-Heeger model.  
In the absence of disorder a domain wall
gives rise to two chiral fractionalized bound states, one at the edge and one bound to the domain wall. On-site disorder
breaks the chiral symmetry, in contrast to hopping disorder.  By using the polarization we find that on-site 
disorder has little effect on 
the chiral nature of the bound states  for weak to moderate disorder.
We explore the behaviour of these bound states for strong disorder, contrasting on-site and hopping disorder 
and connect our results to the localization properties of the bound states and to recent experiments.
\end{abstract}

\maketitle

\section{Introduction}
The Su-Schrieffer-Heeger (SSH) model \cite{Su1979} was introduced in the context
of polyacetylene but has attracted much interest as a model of non-interacting
fermions in one dimension that displays charge fractionalization \cite{Heeger1988}. 
The SSH model also gives a simple example of a model with topologically distinct states which arise for
opposite hopping dimerization patterns.  Fractionalization arises when domain walls
are introduced that separate the two dimerization patterns and give rise to
zero energy modes with specific chiralities bound at the domain walls.

Recently there have been several experimental realizations of the SSH model:
in cold atom systems \cite{Atala2013,Meier2016,Leder2016,Meier2018} and 
graphene nanoribbons \cite{Rizzo2018,Groning2018,Franke2018}.  Condensed matter 
implementations of the SSH model generically break the chiral (sublattice) symmetry
that gives rise to zero modes through e.g. next-nearest-neighbour hopping \cite{PerezGonzalez2019} (as occurs
in polyacetylene) or disorder.  In this work we explore the effect of broken chiral symmetry due to 
disorder on the states that are zero modes in the absence of disorder.  Specifically, this is important for the 
interpretation of experimental results on signatures of chirality in real systems where disorder is inevitable
and chiral symmetry is broken  \cite{Rizzo2018,Groning2018,Franke2018}.

Much of the previous work on disorder in the SSH model has focused on the case in which there 
is disorder in hopping amplitudes \cite{Mondragon-Shem2014,Liu2018,Jurss2019}.  For this special class of disorder, the chiral
symmetry of the model is preserved, and hence zero modes in the clean model remain zero
modes in the disordered model for weak disorder, only disappearing at a critical disorder value \cite{Mondragon-Shem}.  
On-site disorder explicitly breaks chiral symmetry so that
the zero modes in the clean limit are no longer topologically protected and have a non-zero
energy for infinitesimal disorder.  In the limit of infinitesimal disorder we expect there to be states that closely 
resemble the zero energy modes in the clean limit.  Previous work \cite{Munoz2018,PerezGonzalez2019} has investigated how
on-site disorder affects the localization properties of edge states, but not their chiral properties.
In particular, for a SSH model with a 
domain wall, if the system is large enough, we might expect localized states at the wall 
and the edge to retain some of their chiral properties.  

In this work we study the disordered SSH model with a domain wall and explore the extent to which the localized states retain their 
chiral nature even though chiral symmetry has been broken by on-site disorder.
Our main tool to do this is the polarization -- it has been shown that in a system with chiral
symmetry there is a relationship between the winding number and the polarization 
\cite{King-Smith1993,Resta1994,Ortiz1994,Qi2008,Hughes2011,Turner2012,Song2014}.
Our main result is that we find that for even quite sizeable 
disorder strengths, the bound states can be viewed as being chiral from a practical point 
of view, even if not perfectly so.  We relate the changes in polarization as a function of on-site disorder 
strength to changes in the localization properties of the electronic states.

The structure of this paper is as follows: in Sec.~\ref{sec:model} we introduce the disordered SSH model and show 
numerical calculations of its spectrum and of localized states.  In Sec.~\ref{sec:disc} we discuss our 
results and conclude.

\section{The disordered SSH model}
\label{sec:model}
\begin{figure}[ht]
 \includegraphics[width=8cm]{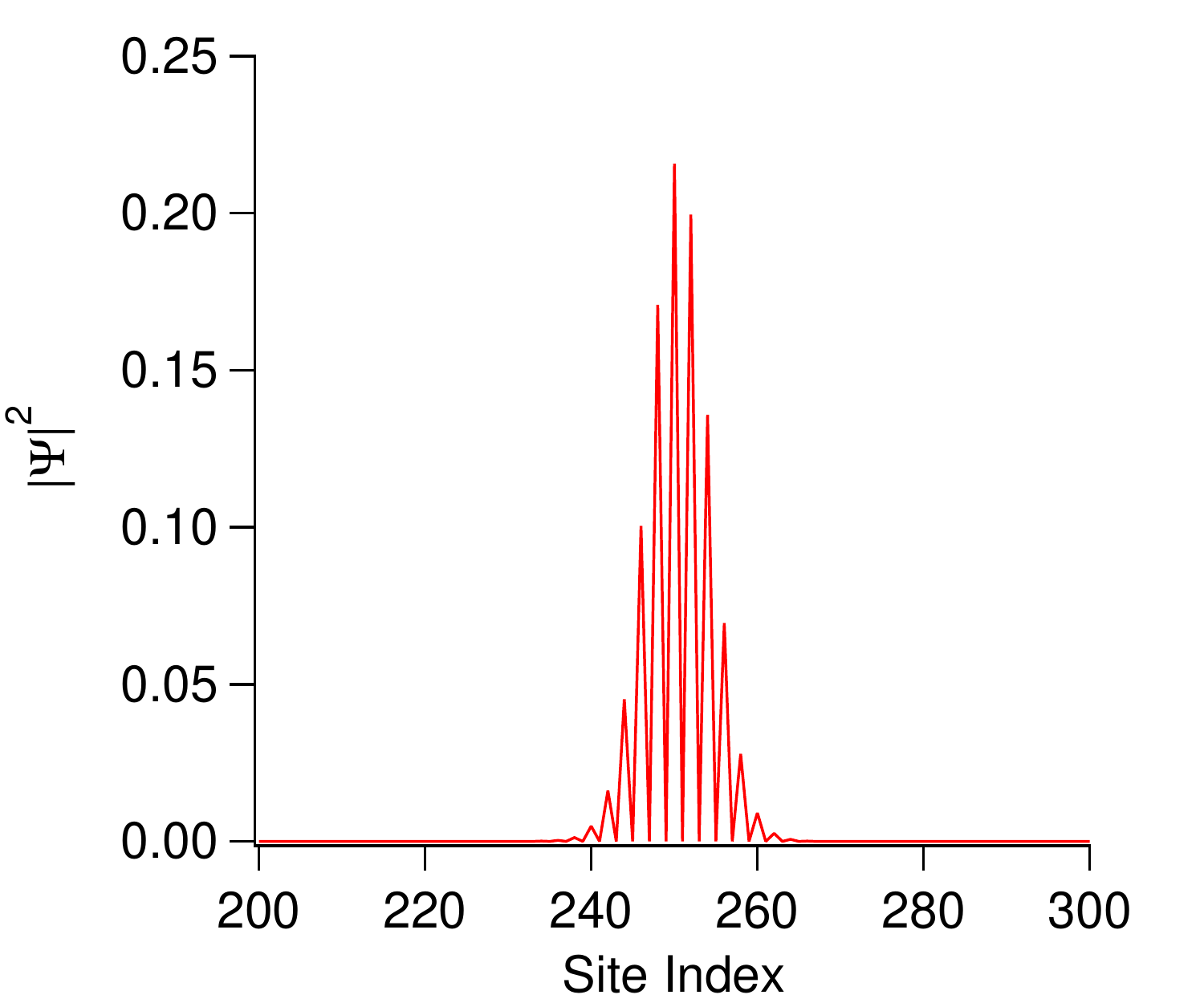}
	\caption{Wavefunction for the zero mode localized on the domain wall in the centre of a $N = 500$ site chain 
in the clean SSH model for $\xi/a = 10$ and $u_0 = 0.2$.}
\label{fig:zero_mode_wavefunctions}
\end{figure}

The Hamiltonian for the SSH model on a $N$ site chain may be written as

\begin{eqnarray}
	H_{\rm SSH} = - t_0 \sum_{n=1}^{N-1} \left[1 + (-1)^n u\right]\left\{c^\dagger_{n+1}c_n + c^\dagger_n c_{n+1}\right\}, 
	\nonumber \\
\end{eqnarray}
where $c_n$ and $c^\dagger_n$ are annihilation and creation operators for fermions on site $n$ respectively, 
$t_0$ is the hopping strength and $u$ is the dimensionless strength of the stagger in the hopping.
Fractionalized states arise if a domain wall is introduced into the parameter $u$ \cite{Su1979,Heeger1988}.  Here we 
consider domain walls of the form:
			\begin{equation}
			u=u_0\tanh\left[\frac{(n-n_0)}{\xi/a}\right],
			\label{eq:dw}
			\end{equation}
where $u_0$ specifies the amplitude of the domain wall, $n_0$ is the centre of the domain wall and 
$\xi$ is the width, with $a$ the lattice spacing.  In the presence of such a domain wall, the 
SSH model develops fractionalized zero modes which have support on a single sublattice.
In a finite chain with open boundary conditions, and a domain wall of the form Eq.~(\ref{eq:dw}),
one of these zero modes will be localized at an edge, and the other will be localized at the domain wall, 
as illustrated in Fig.~\ref{fig:zero_mode_wavefunctions}.  

We introduce disorder in the form of a random on-site potential with Hamiltonian
\begin{equation} 
H_{\rm dis} = \sum_{n=1}^N \epsilon_n c^\dagger_n c_n,
\end{equation}
where $\epsilon_n$ is a random variable drawn from a uniform distribution on $[-W,W]$.  Such a potential 
breaks chiral symmetry and hence the fractionalization seen at $W = 0$ will no longer be present.  However, 
it is still of interest to study how the chirality of the bound states that form  the $W = 0$ zero modes evolve
with increasing disorder.  In particular the question we want to investigate is how they lose their chirality with 
increasing disorder. 

\begin{figure}[ht]
	\includegraphics[width=8cm]{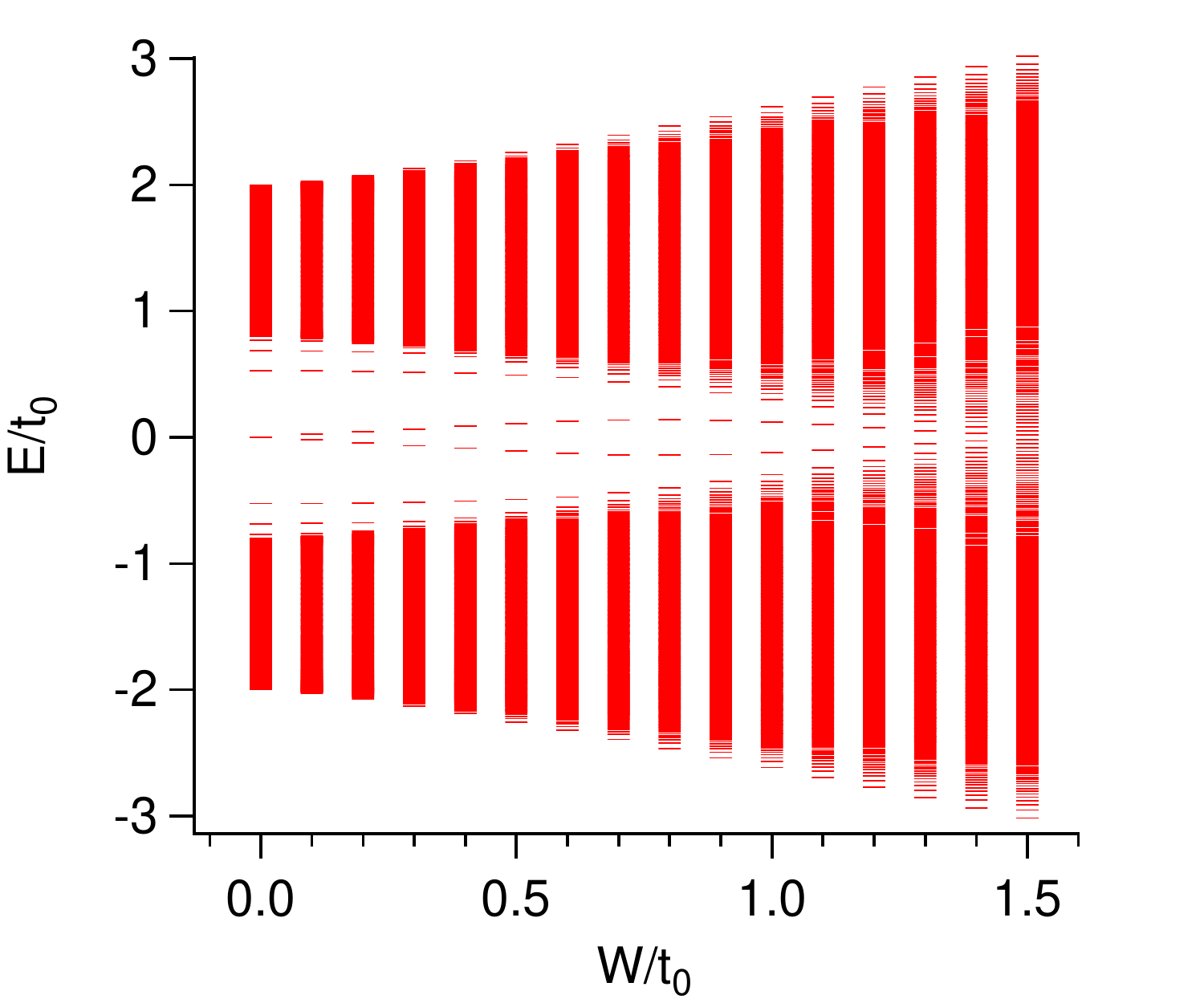}
	\caption{Ordered energy eigenvalues of a $N=500$ site chain in the SSH model  as they evolve with increasing $W/t_0$, for $\xi/a = 10$ and $u_0 = 0.2$}
	\label{fig:dis_spectrum}
\end{figure}

We diagonalized the Hamiltonian $H = H_{\rm SSH} + H_{\rm dis}$ for a $N=500$ site chain and found 
the ordered list of energy eigenvalues.  We averaged over
50000 disorder configurations to obtain Fig.~\ref{fig:dis_spectrum}.  Once chiral symmetry is broken by a disorder
potential ($W \neq 0$), the zero modes seen at $W = 0$ move away from being exactly at zero energy, but they are clearly
identifiable in the gap out to disorder strengths of $W/t_0 \sim 1.3$.  Several other bound states are visible in the gap out to $W/t_0 \sim 0.8$
for the particular choice of parameters in Fig.~\ref{fig:dis_spectrum}.

\begin{figure}[bht]
\includegraphics[width=5.5cm]{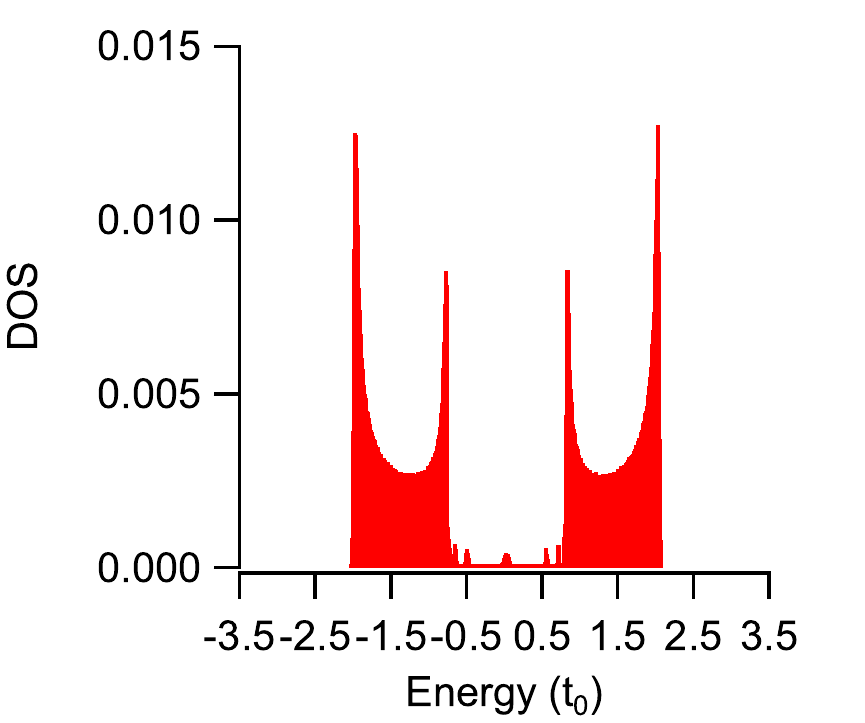}
\includegraphics[width=5.5cm]{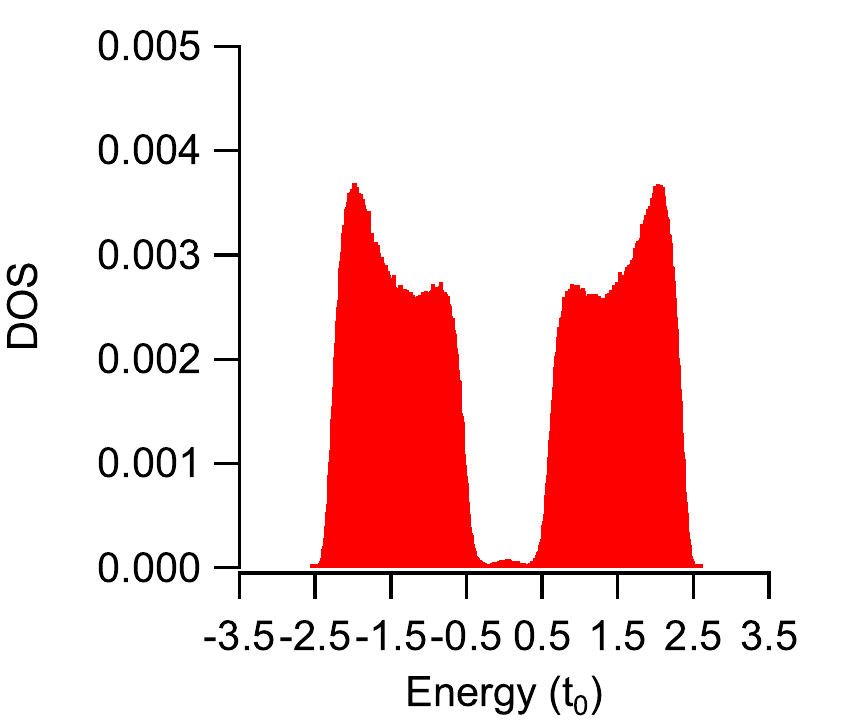}
\includegraphics[width=5.5cm]{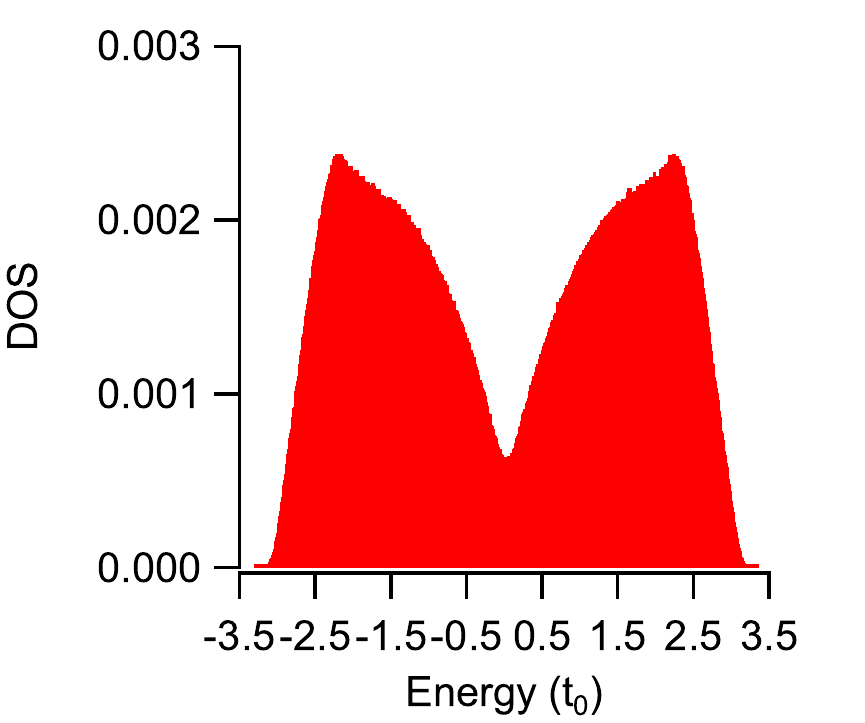}
\caption{Disorder averaged density of states (DOS) for a SSH chain with $N=500$ sites for a domain wall centred 
at site 250 with width $\xi/a=100$, strength $u_0 = 0.2$ and $N = 500$ for disorder strengths 
a) $W/t_0=0.1$, b) $W/t_0=0.7$, and c) $W/t_0=1.5$.}
\label{fig:dos}
\end{figure}

We also calculated the disorder averaged density of states, shown for $W/t_0 = 0.1, 0.7,$ and 1.5 in Fig.~\ref{fig:dos}.
Peaks corresponding to the bound states are clearly visible up to moderate disorder ($W/t_0 \lesssim 1$) but for stronger
disorder the bands broaden sufficiently to obscure them.
While the $W = 0$ zero modes do not continue to have zero energy for $W \neq 0$, we can ask whether they
can be treated as chiral for practical purposes as the disorder is increased.

In the case of disorder that preserves chiral symmetry (e.g. hopping disorder) 
\cite{Mondragon-Shem2014} one can consider
a real-space calculation of a topological invariant which is closely related to the polarization \cite{Song2014}. 
In the case of on-site disorder that we consider here, there is no strict topological
protection, so we instead focus on the polarization of bound states, which can change continuously as disorder increases.  
Specifically, we introduce projection operators $\hat{P}_A$ and $\hat{P}_B$, which project a bound state $\ket{\psi}$ on to 
either the $A$ or $B$ sublattices respectively.  We can use these projection operators to calculate the polarization,
i.e. the density imbalance between $A$ and $B$ sublattices 
		\begin{equation}
			P =\langle\psi|\hat{P}_A-\hat{P}_B|\psi\rangle,
			\label{eq:EOD}
		\end{equation}
for bound states $\ket{\psi}$ localized at the domain wall and the edge.  When $W \neq 0$ we select the bound states by projecting disordered 
bound states $\ket{\psi (W\neq0)}$ onto the $W=0$ bound states $\ket{\phi(W=0)}$.  The results we show are for the states that have the 
maximum overlap with the clean bound states i.e. those that maximize $\left|\left<\psi (W\neq0)|\phi(W=0)\right>\right|$.

\begin{figure}[ht]
	\includegraphics[width=8cm]{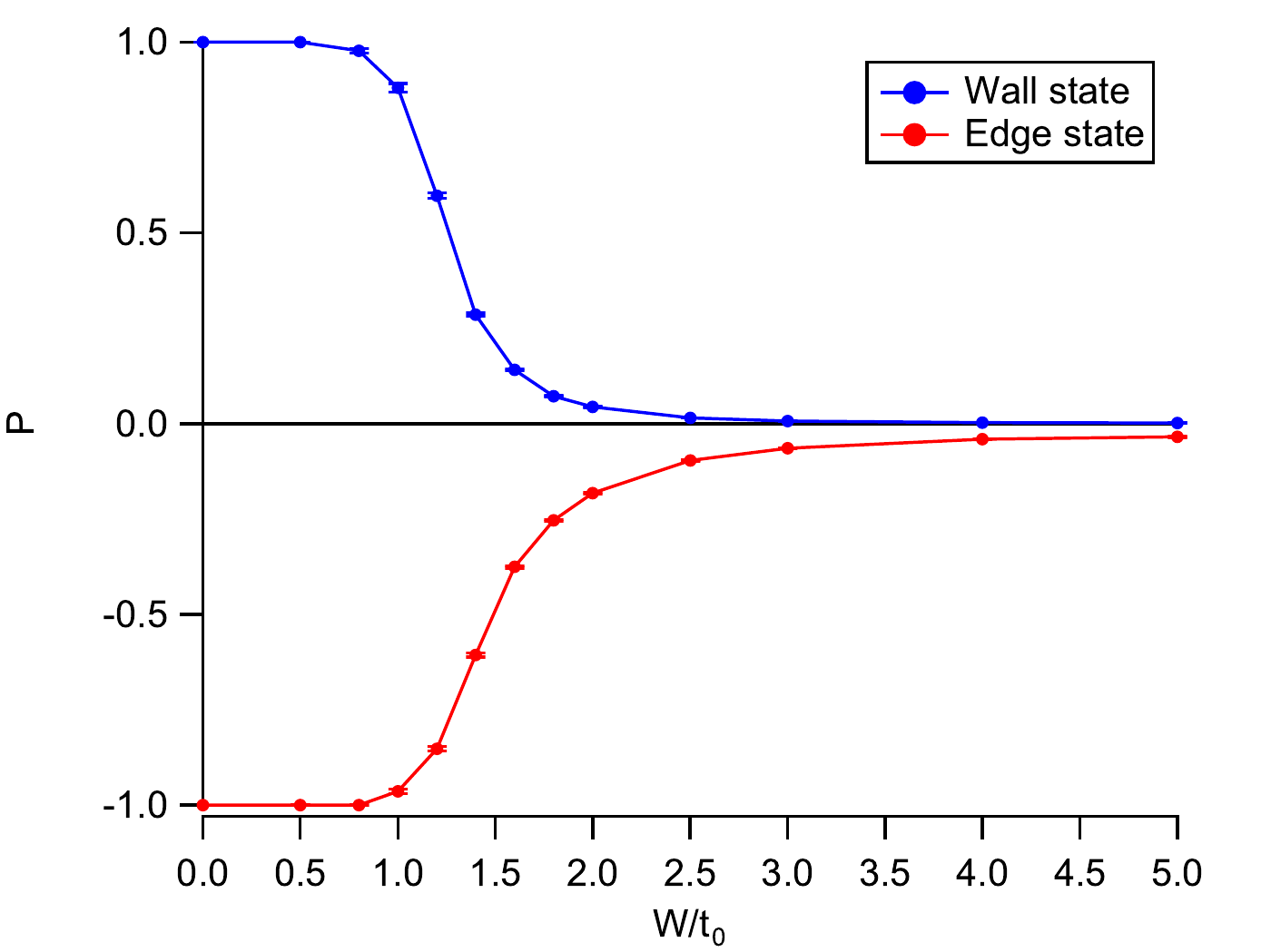}
	\caption{Polarization $P$ for the states bound at the domain wall (Wall state) and the edge (Edge state) in a chain with $N=500$
		        sites with $\xi/a = 10$ and $u_0/t_0=0.2$ for the SSH model with hopping disorder. }
			\label{fig:offd_dis}
		\end{figure}

We calculate the polarization $P$ for these  bound states in the presence of both chiral symmetry preserving and chiral symmetry breaking disorder.  We introduce chiral symmetry preserving disorder via the Hamiltonian
\begin{eqnarray}
	H_{\rm chiral \, dis} = \sum_{n=1}^{N-1}\tau_n\left\{c^\dagger_{n+1}c_n + c^\dagger_n c_{n+1}\right\}, 
	\nonumber \\
\end{eqnarray}
where $\tau_n$ is a random variable drawn from a uniform distribution on $[-W,W]$.
Similarly to Ref.~\cite{Mondragon-Shem2014}, we find that $P$ goes to zero for large $W$ for the $W \neq 0$ bound states at the domain wall and the edge, consistent with the transition in winding
number with $W$ identified in Ref.~\cite{Mondragon-Shem2014}, as illustrated in Fig.~\ref{fig:offd_dis}.  Even though the states
lose their polarization, they remain localized for all disorder strengths \cite{Licciardello1977}.

\begin{figure}[ht]
\includegraphics[width=8cm]{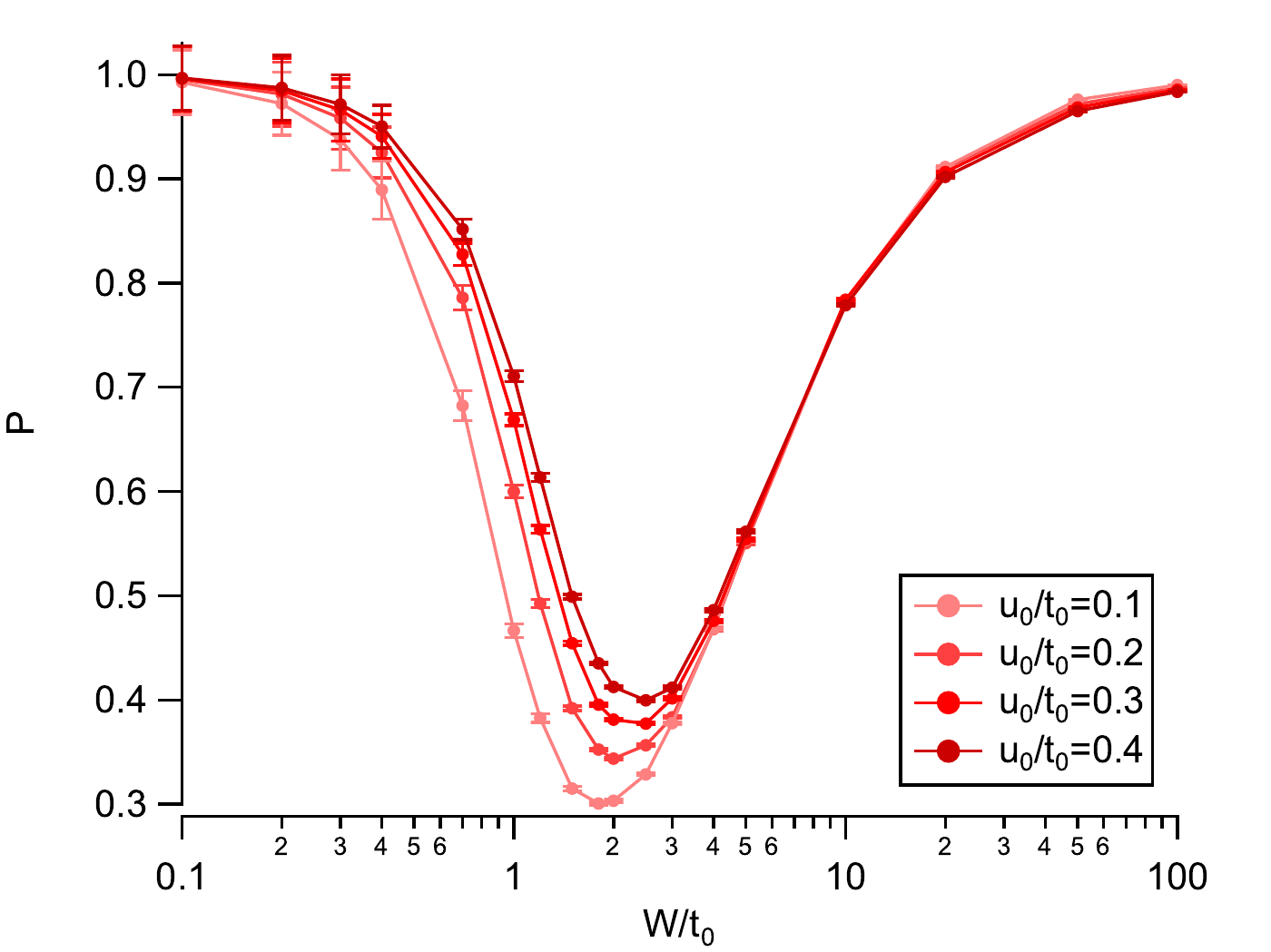}
\includegraphics[width=8cm]{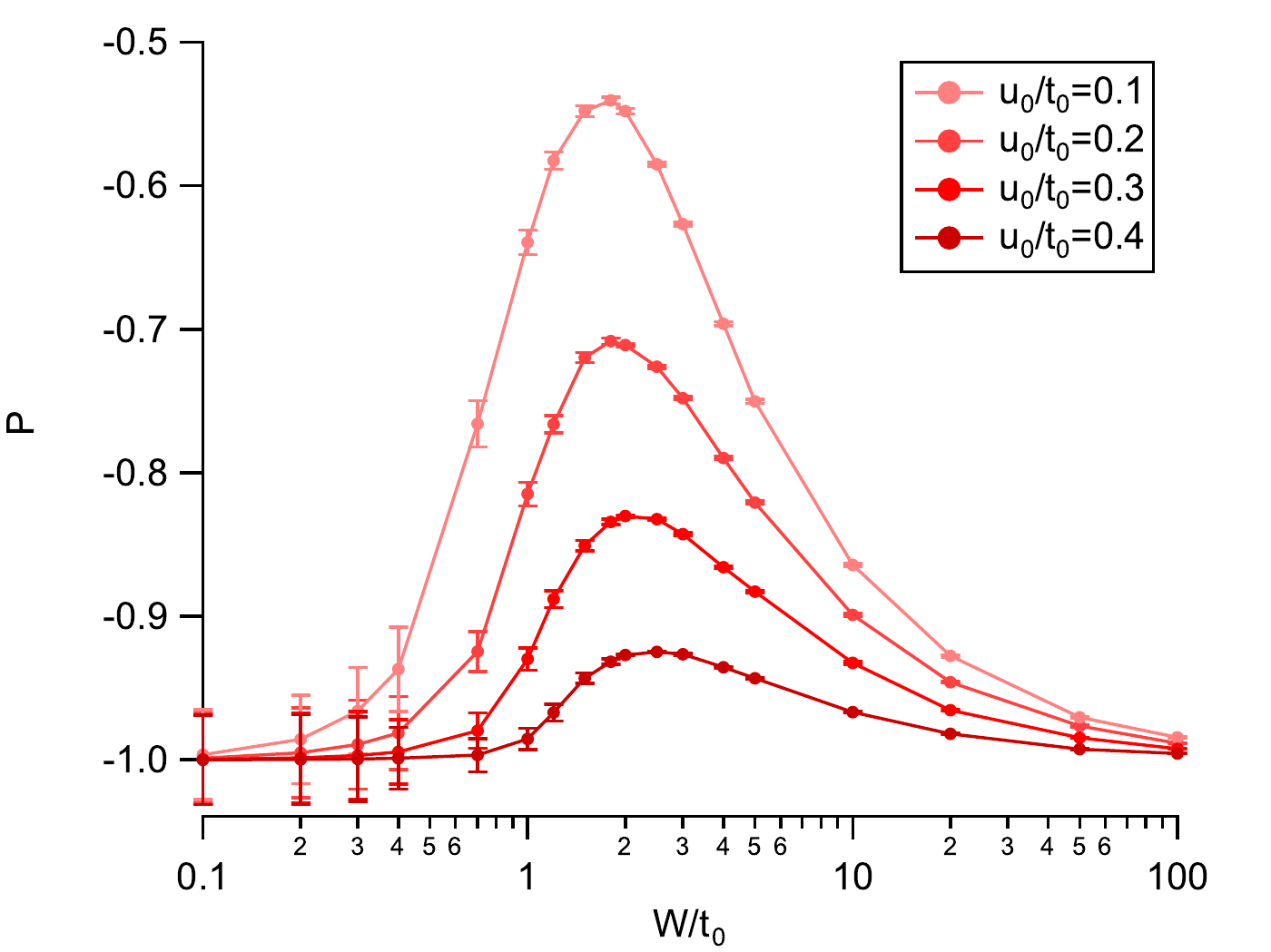}
\caption{Polarizations $P$ for the states bound at the domain wall and the edge in a chain with 500 sites and domain wall width $\xi/a = 10$ for the SSH model with on site disorder 
 for four different domain wall strengths  $u_0/t_0 = 0.1, 0.2, 0.3,$ and 0.4.}
\label{fig:diag_dis}
\end{figure}

We performed similar calculations for the SSH model with on-site disorder and display the results in Fig.~\ref{fig:diag_dis}.
We found that for $N \gtrsim 300$ that our results appear to be 
independent of $N$.  We also see that $|P|$ decays more quickly with $W/t_0$ than 
for hopping disorder, but that $1-|P| \ll 1$ for values of $W/t_0$ that are an appreciable fraction of 1, demonstrating that small amounts
of on-site disorder do not greatly alter the chiral nature of the states.

\begin{figure}[H]
\includegraphics[width=8cm]{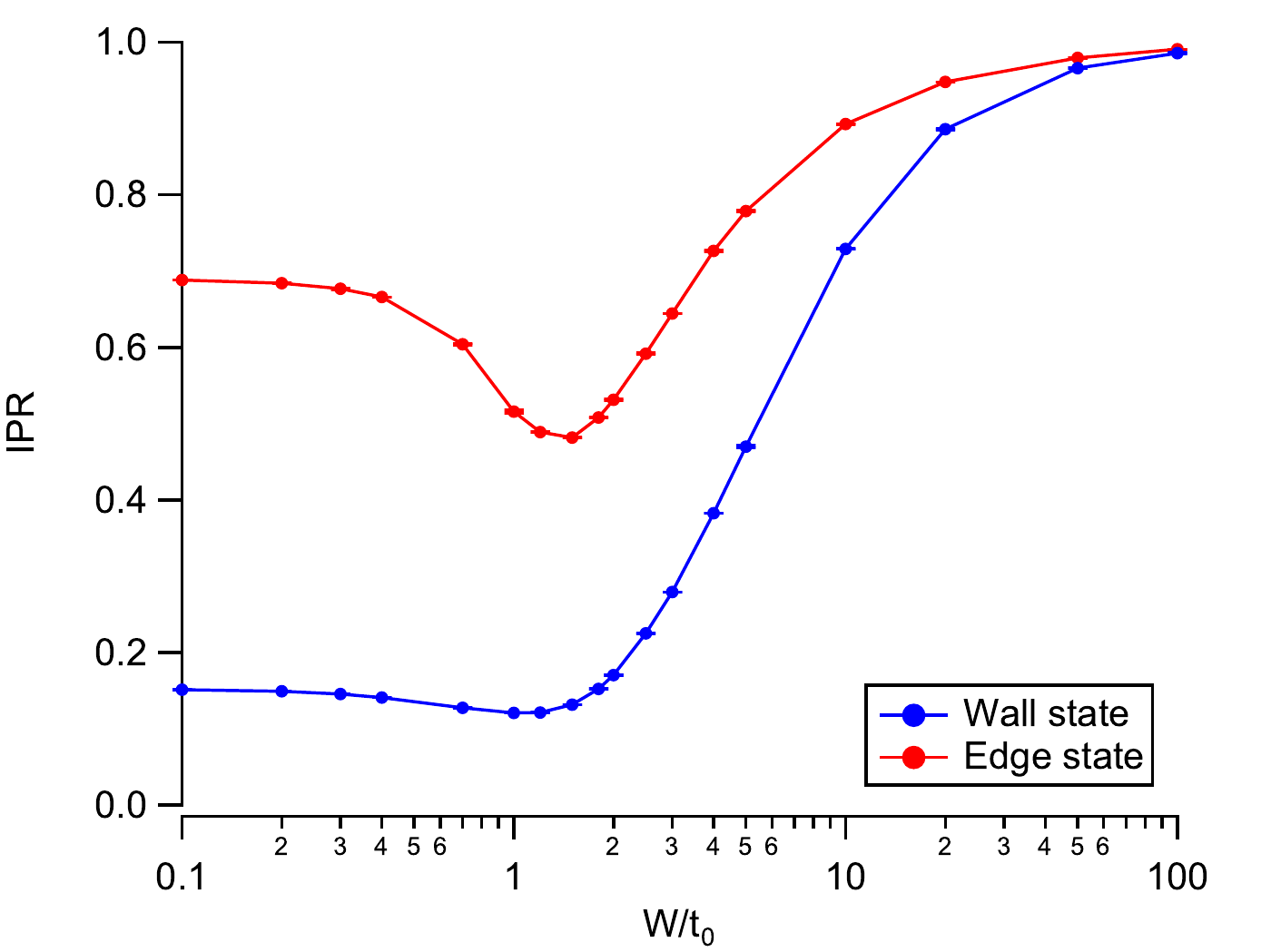}
\caption{Inverse Participation Ratio (IPR) for the states bound at the domain wall (Wall state) and the edge (Edge state) 
	in a chain  with 500 sites and a 
 domain wall width of $\xi/a = 10$ and $u_0=0.2$ as a function of disorder strength.}
\label{fig:IPR}
\end{figure}

Unlike the situation in which there is hopping disorder, $|P|$ does not approach zero with increasing 
$W/t_0$ and in fact increases towards 1 with increasing $W/t_0$.  The reason for this behaviour can be illuminated with the 
inverse participation ratio, defined by 
\begin{equation}
	{\rm IPR} = \frac{\sum_i \left|\psi(\bvec{r}_i)\right|^4}{\left|\sum_i \left|\psi(\bvec{r}_i)\right|^2\right|^2},
\end{equation}
which gives a measure of localization. The value of the IPR differs significantly between localized and extended states.
For localized states, the IPR takes a constant value, whereas for extended states, the IPR scales like $1/L^d$ where $d$ is the 
spatial dimension.  The IPR is illustrated in Fig.~\ref{fig:IPR} and illustrates that the localization length increases up 
to a disorder strength of $W/t_0 \sim 1-2$, consistent with results obtained for edge states in smaller 
systems \cite{Munoz2018,PerezGonzalez2019}.
The localization length decreases at larger values of disorder, consistent with Anderson localization
becoming more important.  The states are always localized, as expected for a one dimensional disordered fermion system \cite{Mott1961}, but 
the degree of localization varies with disorder strength.

The behaviour seen in $P$ can be understood from a picture in which increasing on-site disorder
breaks chiral symmetry so that the zero disorder zero mode states start to have some support on both sublattices, but unlike the hopping
disorder case, $P$ does not go to zero, because with increasing on-site disorder strength, the states become sufficiently localized
that most of their support is on a single site.  Figure~\ref{fig:IPR} illustrates that there is a crossover from a localized state that
retains much of its $W = 0$ chiral character to a strongly Anderson localized state as a function of $W/t_0$.

\section{Discussion}
\label{sec:disc}
We studied the SSH model with on-site disorder and compare our results to those obtained for hopping disorder.
Our results demonstrate that even though chiral symmetry is broken by the introduction of on-site disorder, the zero 
energy states at zero disorder evolve so that they continue to be strongly polarized for $W/t_0 \lesssim 1$
and can be treated as chiral for practical purposes for moderate on-site disorder.  The IPR illustrates that this 
is a crossover from topology-induced localization to Anderson localization with increasing disorder. 

We note that our calculations here have direct relevance to recent experiments.  In particular, two groups used graphene nanoribbons
\cite{Rizzo2018,Groning2018,Franke2018}
to engineer the SSH model and studied edge states in these systems.  Our results here show that the edge states that are topologically
protected in the clean limit persist to large values of disorder. Hence, given the inevitability of some level of on-site disorder
in experiment, the edge states observed in experiment are still meaningful approximations to the clean case.
From a theoretical perspective, the fractionalization \cite{Jackiw1976} seen in the 
SSH model in one dimension has been generalized to 
two dimensions \cite{Hou2007,Chamon2008,Seradjeh2008,Roy2012} and it would be
very interesting to see how disorder affects the zero energy modes in those models.

\section*{Acknowledgements}
M. P. K. was supported by NSERC.

\end{document}